\documentclass[10pt,aip,graphicx,amssymb,amsmath,superscriptaddress,twocolumn]{revtex4-1}

\usepackage{graphicx}
\usepackage{dsfont}
\usepackage{bbold}
\usepackage{bm}

\begin{document}


\title{Analytic model of effective screened Coulomb interactions in a multilayer system} 



\author{H. Ouerdane}
\affiliation{Mediterranean Institute of Fundamental Physics, Via Appia Nuova 31, 00040 Marino, Rome, Italy}
\affiliation{CNRT Mat\'eriaux UMS CNRS 3318, 6 Boulevard Mar\'echal Juin, 14050 Caen Cedex, France}

\date{\today}

\begin{abstract}
The main objective of the present work is the development of an analytically tractable model of screened electron-electron and electron-exciton interactions in layered systems composed of two parallel semiconductor quantum wells separated by a dielectric layer. These systems are promising for superconductivity with excitons-polaritons, and spin manipulation. Polarization effects induced by the dielectric mismatch in the nanostructure are taken into account using the image charge method. The obtained analytic expressions are used to calculate screened electron-electron and electron-exciton interactions; these are compared to results computed using other recently published models.
\end{abstract}


\maketitle 

\section{Introduction}
In condensed matter theory, the exact calculation of interactions and physical observables on the analytical level are possible only for a limited number of situations despite considerable advances of theoretical methods \cite{Fetter,Negele}. Exactly solvable models of non-trivial quantum many-body problems (see, e.g., Refs. \cite{Hewson,Eckle,Ouerdane07,Ouerdane08} and references therein) are useful for a variety of purposes, but their scope remains restricted. In fact, only approximate results can be obtained for most cases of interest owing to the complex interplay between interparticle interactions and quantum statistics.

In solid-state systems, screening of the Coulomb interaction between electrically charged quasiparticles is one of the main manifestations of the many-body correlations. An important consequence of Coulomb screening is indeed to alter the long-range nature of the interaction in such way that it acquires a short-range character. Physical quantities of interest are the polarization function and the related dielectric function from which various observables of experimental relevance, describing either the local or the global response of a system to an external perturbation, can be derived. 

A weakly interacting electron gas in a semiconductor sample constitutes one of the simplest model systems. The \emph{effective} screened interaction potential experienced by a test charge in an electron gas reads

\begin{equation}\label{veff}
V_{\rm eff}(\bm{q},\omega)=\frac{V(\bm{q})}{\varepsilon(\bm{q},\omega)},
\end{equation}

\noindent where $V(\bm{q})$ is the Fourier transform of the unscreened Coulomb potential of the test charge, and $\varepsilon(\bm{q},\omega)$ is the dielectic response function given by the well-known Lindhard formula \cite{LindhardJ,Koch}:

\begin{equation}\label{Lindhard}
\varepsilon(\bm{q},\omega)=1-V(\bm{q})
\sum_{\bm{k}}\frac{f_{\bm{k}-\bm{q}}-f_{\bm{k}}}{\hbar(\omega+i\delta)+E_{\bm{k}-\bm{q}}-E_{\bm{k}}},
\end{equation}

\noindent where $f_{\bm{k}}$ is the distribution function of electrons in momentum space, $E_{\bm{k}}$ is the dispersion, and $\bm q$ is the transferred wavevector. The presence of the small frequency parameter $\delta$ ensures an adiabatic switch-on of the test charge potential. Formula \eqref{Lindhard}, obtained from the averaged screened density operator in the random phase approximation fed to the Poisson equation, is valid for both equilibrium and nonequilibrium situations; it can be considered as the starting point of screened interaction calculations in various systems. In this article, the systems under consideration are in quasi-thermal equilibrium.

The case of a semiconductor electron-hole plasma \cite{Zimmermann}, i.e. a two-component system, represents the next level of complexity. One of the first attempts to treat such case consisted of introducing a variational parameter to characterize the total plasma screening in order to compute numerically the ground state of an electron-hole pair \cite{Lee,Spector}.
A multiband plasmon-pole approximation was used in Ref.~\cite{Pereira1} to study nonlinear optical properties of coupled-band semiconductor quantum wells. More recently the total plasma screening was defined as the sum of the separate contributions of each carrier gas in studies of the statistical mechanics and ionization degree of the mixed exciton/electron-hole gas in quasi-equilibrium \cite{Portnoi1,Portnoi2}, and ultrafast relaxation of nonequilibrium spin-polarized electron-hole plasma \cite{Ouerdane1}. A more rigorous approach to plasma screening in mixed exciton/electron-hole systems has to account for the interaction between the different plasma species. This was achieved in the particular problem of the Mott transition in quasi-two-dimensional semiconductor systems including screening by spatially indirect excitons \cite{Nikolaev1,Nikolaev2}.

Semiconductor systems which confine the carrier gases along one spatial dimension emerged several decades ago and have been extensively studied since then; many of the main properties of two-dimensional systems, of which Coulomb screening, are presented and discussed in the thorough review of Ando \emph{et al}~\cite{Ando}. Systems composed of two parallel quantum wells facing each other exhibit particular interactions between the separated electron gases such as the so-called frictional Coulomb drag\cite{Gramila1,Gramila2}. These interwell interactions were theoretically studied in detail through the calculation of the temperature-dependent rate of momentum transfer in the framework of Boltzmann equation\cite{Jauho1} and Kubo linear response theory\cite{Jauho2,Kamenev}. Experimentally, the electron-electron interactions and related momentum transfer rates are probed by the drag-induced voltage\cite{Gramila3}.

Nanostructures built from layers with varying dielectric properties continue to present a great interest both from fundamental and applied viewpoints. Recently an original idea was put forward to possibly observe high-temperature superconductivity: in a sandwiched structure consisting of an n-doped quantum well placed between a pair of parallel quantum wells made of different materials, embedded in a microcavity, the excitations of an exciton-polariton Bose-Einstein condensate serve as a binding agent of Cooper pairs \cite{Laussy1,Laussy2}. In such structure, it was also shown numerically that due to interactions in the hybrid Bose-Fermi system a roton minimum may appear in the spectrum of elementary excitations of the exciton condensate \cite{Roton}. In the context of spintronics, screened electron-electron interactions in multiple quantum wells systems are of great importance for practical applications such as spin manipulation. Indeed these screened interactions are a source of a time-dependent randomness in the spin-orbit coupling, and Coulomb screening enhances the spin relaxation time; see Refs.~\cite{Glazov1,Glazov2,Glazov3} for further details.

In these works, the model screened Coulomb interactions account for the interwell distances, but not for the effects of the \emph{dielectric mismatch} in the layered structures. Neglect of this key aspect of the electrostatic problem surely has an impact on the amplitudes of the computed interaction matrix elements. In addition, if systems are characterized by an important dielectric mismatch, this may possibly affect some of the conclusions of the works.  The development of a model of screened interactions in such systems thus is a topical problem and should also prove useful to compare and assess the validity of the previously employed screening models.

In the present work, a generic system composed of two parallel semiconductor quantum wells separated by a dielectric layer is considered; each layer is characterized by a dielectric constant. The main purpose is to obtain a \emph{tractable} model of plasma screening that can reasonably describe the following interactions: \emph{i}/ `electron-electron' within the same quantum wells (intrawell interaction), \emph{ii}/ `electron-electron', each being part of a gas residing in one of the two separate quantum wells (interwell interaction); and \emph{iii}/ `electron-exciton', with the electron and exciton gases being in one of the two separate quantum wells (interwell interaction). The calculation of these interactions will take into account the dielectric mismatches at the layers interfaces.

An exact analytic solution to such kind of electrostatic problems is obviously out of reach, and a full computational treatment based on the numerical integration of the Poisson equation with suitable boundary conditions at the interfaces, also highly desirable, would need to be tested against analytical results derived from simplified models. Therefore, a realistic goal is to obtain a description of the processes at work and estimates for the amplitudes of the interaction matrix elements considering a simple approach involving standard arguments from classical electrodynamics and quantum mechanics.

The article is organized as follows: in Section II, brief reviews of the model of statically screened Coulomb interactions in an ideal quantum well, and the method of image charges to account for the influence of the polarization effects induced by the dielectric mismatch, are given; section III is devoted to the derivations of the analytical formulas for the screened (intra- and interwell) electron-electron interactions; the interaction of an electron gas in a quantum well with a neutral exciton gas in a separate parallel quantum well is the object of section IV. The obtained interactions are compared with the published ones and discussed.

\section{General considerations}

\begin {figure}[!rh]
\centering
\scalebox{.325}{\rotatebox{0}{\includegraphics{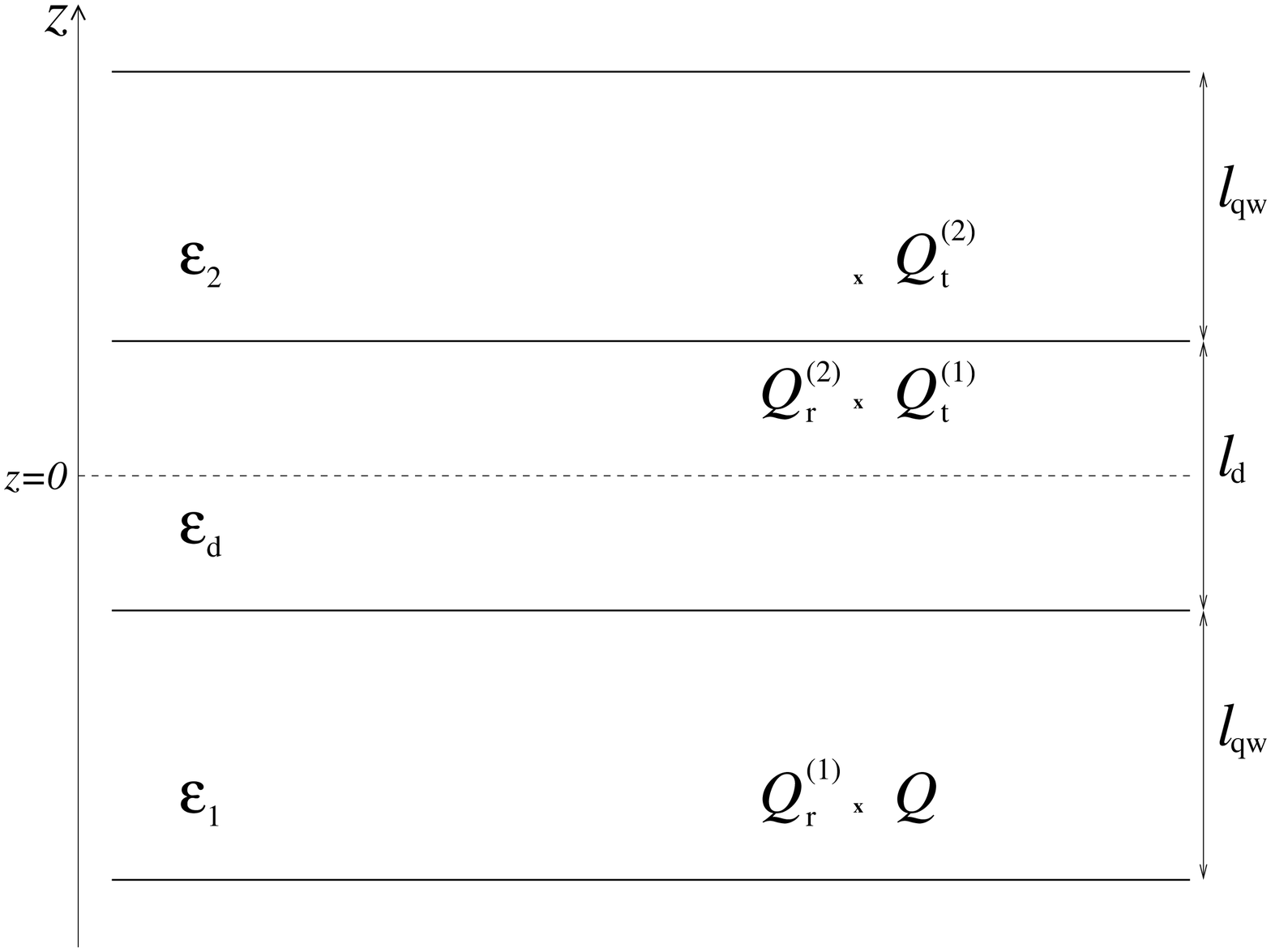}}}
\caption{Charge $Q$ (located in the quantum well 1) and its first and second order transmitted and reflected image charges (labeled with the indices {\rm t} and {\rm r} respectively). See text. The parameters $\varepsilon_1$, $\varepsilon_{\rm d}$, and $\varepsilon_2$ are the relative static permittivities of each medium.}\label{stack}
\end {figure}

Modulation-doped semiconductor quantum wells are considered. The doping is assumed to be sufficiently small so that the dielectric constants of the barriers materials do not differ much from those of the wells (such systems are routinely designed, taking, e.g., GaAs as the well material and AlGaAs as the barrier material). The model system considered for the present work is composed of two such quantum wells, labeled 1 and 2, which are parallel and separated by a dielectric layer of thickness $l_{\rm d}$ and dielectric constant $\varepsilon_{\rm d}$, as depicted in Fig.~\ref{stack}. In this case, it is reasonnable to assimilate the two quantum wells to two semiconducting layers characterized by two dielectric constants $\varepsilon_1$ and $\varepsilon_2$ and two electron effective masses are $m_1$ and $m_2$. For simplicity, a symmetric structure is assumed: the thicknesses of the wells are equal: $l_{\rm qw1}=l_{\rm qw2}=l_{\rm d}=l_{\rm qw}=2L$.

To illustrate the results obtained in this work, numerical calculations will be performed assuming that the system is simply composed of two parallel GaAs-based modulation-doped semiconductor quantum wells separated by a vacuum layer. The following parameters will be taken: $\varepsilon_1 = \varepsilon_2 = 13.71$, $\varepsilon_{\rm d} = 1$, $m_{\rm e} = 0.0665 m_0$, and $m_{\rm h} = 0.457 m_0$ for the electron and heavy hole effective masses, where $m_0$ is the free electron mass. The electron densities will be taken as $10^{12}~\mbox{cm}^{-2}$ and two temperature regimes will be considered: room temperature ($T$ = 300 K) and cryogenic temperature ($T$ = 4 K).

\subsection{Screening in a two-dimensional electron gas}
\label{scrn}

Consider a two-dimensional electron gas of density $N$ at temperature $T$, perfectly confined in a semiconductor quantum well with a static relative permittivity $\varepsilon$. The electron effective mass is $m$. The full RPA dielectric function \eqref{Lindhard} exhibits a continuum of poles. As a consequence the numerical evaluation of the screened Coulomb potential \eqref{veff} is impractical. To circumvent this problem, it is possible to modify this formula by making further approximations. The simplest approach \cite{Koch} is to consider both the limits of static screening, $\omega\rightarrow 0$, and long wavelength, $q\rightarrow 0$, and to define an effective plasmon pole. In these conditions, the inverse of the 2D dielectric function can be written as follows \cite{Koch}:

\begin{equation} \label{eqA_27}
  \frac{1}{\varepsilon(\bm q)} = \frac{q}{q+\kappa},
\end{equation}

\noindent where $\kappa$ is the screening parameter:

\begin{equation}\label{kappa0}
\kappa = \frac{\displaystyle m e^2}{\displaystyle 2\pi\varepsilon_0\varepsilon \hbar^2}~ \left(1 - e^{-\hbar^2\beta\pi N/m}\right),
\end {equation}

\noindent and $\beta = 1/k_{\rm B}T$ is the inverse thermal energy. Practical applications show that it is often sufficient to use the much simpler static plasmon-pole approximation instead of the full RPA dielectric function to obtain good qualitative results.

\subsection{Coulomb potential: the method of images}
\label{imgchrg}

As can be seen in Fig. \ref{stack}, the main difficulty to determine the screened Coulomb potentials in the quantum wells arises from the discrete layered structure of the system characterized by abrupt changes of the dielectric constants at the interfaces. A dielectric mismatch always induces polarization effects that must be accounted for. For instance, the role of dielectric mismatch has been evidenced in the obervation of the reduction in the charging energy of the shallow state of a negatively charged dopant in silicon\cite{Calderon}. The method of images from classical electrostatics \cite{Griffiths} is very well suited for such type of problems and it has been succesfully adapted in the framework of the GW approach to the study of the two-dimensional dielectric screening in systems formed of repeated layers of different dielectric materials \cite{Freysoldt}.

The method of images is best explained following the standard textbook case \cite{Griffiths} of the two semi-infinite dielectric media, each characterized by a dielectric constant, $\varepsilon_1$ and $\varepsilon_2$ respectively, and separated by a plane interface. Assuming the presence of an electric charge $Q$ located at ${\bm r}_Q$ in the medium 1, the electrostatic potential at ${\bm r}$ is 

\begin{equation}\label{pot1}
V_1 = \frac{1}{4\pi\varepsilon_0\varepsilon_1}~\left(\frac{Q}{|{\bm r} - {\bm r}_Q|} + \frac{Q'}{|{\bm r} - {\bm r}'|}\right),
\end{equation}

\noindent if ${\bm r}$ is in the medium 1, in which case ${\bm r}'$ is the image of ${\bm r}$ through the interface; and

\begin{equation}\label{pot2}
V_2 = \frac{1}{4\pi\varepsilon_0\varepsilon_2}~\frac{Q''}{|{\bm r} - {\bm r}_Q|},
\end{equation}

\noindent if ${\bm r}$ is in the medium 2. Imposing continuity at the interface for both the electric field and electrostatic potential yields:

\begin{subequations}
\begin{equation}\label{qtrans}
Q'=\frac{\varepsilon_1 - \varepsilon_2}{\varepsilon_1 + \varepsilon_2}~Q~~~~~\mbox{``transmitted''}
\end{equation}
\begin{equation}\label{qrefl}
Q''=\frac{2\varepsilon_2}{\varepsilon_1 + \varepsilon_2}~Q~~~~~\mbox{``reflected''}
\end{equation}
\end{subequations}

\noindent The charge images are thus of two kinds, the ``transmitted'' image charge, $Q'$, located at the point opposite to that of $Q$ with respect to the interface plane, and the ``reflected'' image charge, $Q''$, which is located at the same point as $Q$. These image charges are merely convenient mathematical objects used to compute the electrostatic potentials and have no physical reality. The justification of this method comes from the \emph{uniqueness theorem} \cite{Griffiths}, which states that for a given set of boundary conditions there can only be one unique solution to the electrostatic problem. Therefore the potentials \eqref{pot1} and \eqref{pot2} computed with $Q'$ and $Q''$, \eqref{qtrans} and \eqref{qrefl} respectively, are the correct ones.

\section{Screened electron-electron interactions in a multilayer system}

\subsection{Intrawell electron-electron interaction}
\label{eeintsec}

Let us now turn to the structure in Fig. \ref{stack}. Following the analysis above, the first and second order image charges are obtained in a fashion similar to the iterative scheme of Ref.~\cite{Freysoldt}, and the potentials acting within each of the separate quantum wells $i=1$ and $i=2$ are given by: 

\begin{equation}\label{uiipot}
U_{ii} = \frac{e}{4\pi\varepsilon_0\varepsilon_i}~\left[\frac{1}{r_{0i}}+\frac{\alpha_{{\rm t}i}^{(1)}+\alpha_{{\rm r}i}^{(2)}}{r_{1i}} + \frac{\alpha_{{\rm t}i}^{(2)}}{r_{2i}}\right],
\end{equation}

\noindent which is the unscreened potential in the quantum well $i = $ 1 or 2, created by an electron of this well, with an effective mass $m_i$, a charge $e$, and located in $(r_{i\parallel};z_{0i})$. The first and second order image charges parameters in \eqref{uiipot} are defined as

\begin{equation}\label{qt1st}
\alpha_{{\rm t}i}^{(1)}=\frac{\varepsilon_i - \varepsilon_{\rm d}}{\varepsilon_i + \varepsilon_{\rm d}}
\end{equation}

\noindent and

\begin{subequations}
\begin{equation}\label{qt2nd}
\alpha_{{\rm t}i}^{(2)}=\frac{\varepsilon_{\rm d} - \varepsilon_j}{\varepsilon_{\rm d} + \varepsilon_j} ~\frac{\varepsilon_i - \varepsilon_{\rm d}}{\varepsilon_i + \varepsilon_{\rm d}}
\end{equation}
\begin{equation}\label{qr2nd}
\alpha_{{\rm r}i}^{(2)}=\frac{2\varepsilon_j}{\varepsilon_{\rm d} + \varepsilon_j}~\frac{\varepsilon_i - \varepsilon_{\rm d}}{\varepsilon_i + \varepsilon_{\rm d}}
\end{equation}
\end{subequations}

\noindent where $j\neq i$. Note that depending on the values of dielectric constants involved in the above defintions of the $\alpha$ parameters, these may be negative. The distances $r_{\ell i}$ ($i=1,2$ and $\ell=0,1,2$), are defined as follows:

\begin{equation}
r_{\ell i} = \sqrt{r_{i\parallel}^2 + (z-\gamma_{\ell i})^2}
\end{equation}

\noindent where $\gamma_{01} = z_{01}$ ($z_{01}<0$ and $|z_{01}|>L$), $\gamma_{11} = -2L - z_{01}$, $\gamma_{21} = 4L+z_{01}$, $\gamma_{02} = z_{02}$ ($z_{02}>L$), $\gamma_{12} = z_{02} - 2L$, $\gamma_{22} = -4L +z_{02}$. The origin of the $z$-axis, $z=0$, is located half way through the dielectric layer, in the plane parallel to the interfaces. The coordinate $z_{01}$ is in the interval $[-L-l_{\rm qw};-L]$, and $z_{02}$ is in the interval $[L;L+l_{\rm qw}]$ with $l_{\rm qw}=2L$.

To proceed with the calculations on the analytical level, we make further assumptions. The first one is to neglect (for the moment) the $z$-dependence of the potential.

\subsubsection{Ideal confinement} In this case: $r_{0i}=r_{1i}=r_{2i}=r_{i\parallel}$, and the potential reduces to:

\begin{equation}\label{uii}
U_{ii} = \frac{e}{4\pi\varepsilon_0\varepsilon_{{\rm eff}i}}\times \frac{1}{r_{i\parallel}}.
\end{equation}

\noindent where $\varepsilon_{{\rm eff}i}$ is a quantity defined as the effective relative static permittivity:

\begin{equation}
\varepsilon_{{\rm eff}i} = \frac{\varepsilon_i}{1 + \alpha_{{\rm t}i}^{(1)}+\alpha_{{\rm r}i}^{(2)}+\alpha_{{\rm t}i}^{(2)}}.
\end{equation}

Using \eqref{uii}, the derivation of an effective screening parameter is immediate \cite{Koch}:

\begin {equation} \label{kappa}
\kappa_{{\rm eff}i} = \frac{\displaystyle m_ie^2}{\displaystyle 2\pi\varepsilon_0\varepsilon_{{\rm eff}i} \hbar^2}~ \left(1 - e^{-\hbar^2\beta\pi N_i/m_i}\right)
\end {equation}

\noindent where $N_i$ is the 2D carrier concentration in the quantum well $i$.

\begin {figure}[!rh]
\centering
\scalebox{.325}{\rotatebox{0}{\includegraphics*{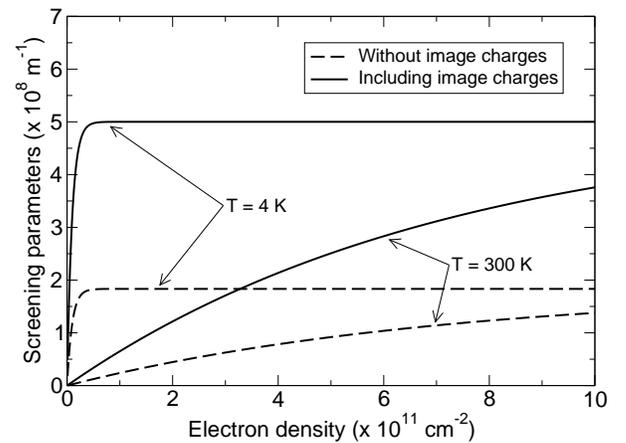}}}
\caption{Effect of image charges on the two-dimensional computed screening parameters. The observed enhancement induced by the presence of the image charges is important in the two temperature regimes.}\label{kappac}
\end {figure}

For a given temperature of the electron gas, the observed amplitude difference between $\kappa_{{\rm eff}i}$ \eqref{kappa} and $\kappa$ \eqref{kappa0} in Fig.~\ref{kappac}, results from the renormalized value of the dielectric constant in the layer $i$: in this example, $\varepsilon_i = 13.71$ and $\varepsilon_{{\rm eff}i} = 5.03$. In the limit $N_i \rightarrow 0$, $\lim_{N_i\rightarrow 0} \kappa_{{\rm eff}i} = \lim_{N_i\rightarrow 0} \kappa = 0$. It is also important to note here that if one sets $\varepsilon_1=\varepsilon_2=\varepsilon_{\rm d}$, then $\varepsilon_{{\rm eff}i} = \varepsilon_i$, and one recovers the standard result for the two-dimensional screening parameter.

The Fourier transform of the electron-electron interaction in the quantum well $i$ reads:

\begin{equation}\label{vii}
V_{ii}({\bm q}) = \frac{e^2}{2\varepsilon_0\varepsilon_{{\rm eff}i}{\mathcal A}}~\frac{1}{q+\kappa_{{\rm eff}i}},
\end{equation}

\noindent where ${\mathcal A}$ is the surface area of the quantum well $i$.

\subsubsection{Accounting for the thickness of the layers}

The simplest approach is to calculate first the partial Fourier transform of the potential $U_{ii}({\bm r}_{i\parallel},z)$, i.e., the Fourier transform restricted to the integration over the ${\bm r}_{i\parallel}$ coordinate, and then perform an average over the well thickness $l_{\rm qw}$ accounting for the transverse component of the single-electron wavefunction: $\psi(z) = \sqrt{2/l_{\rm qw}} \times \sin (\pi z/l_{\rm qw})$. The last step is to add in an \emph{ad hoc} fashion the effective screening parameter \eqref{kappa}, which contains the contributions of the electronic charge $e$ and its images.

Since $r_{0i}$, $r_{1i}$, $r_{2i}$, and $r_{i\parallel}$ are now all different, the electrostatic potential \eqref{uiipot} reads as the sum of three contributions:

\begin{equation}
\nonumber
U_{ii}({\bm r_{i\parallel}},z) = U_{ii}^{(0)}+U_{ii}^{(1)}+U_{ii}^{(2)},
\end{equation}

\noindent where the definition of the three contributions $U_{ii}^{(\ell)}$, with $\ell=0,1,2$, is evident: $U_{ii}^{(0)}$ is the contribution of the physical charge $e$; $U_{ii}^{(1)}$ is the sum of the contributions of the first order transmitted and second order reflected image charges; $U_{ii}^{(2)}$ is the contribution of the second order transmitted image charge. The partial Fourier transforms of these contributions are of the form:

\begin{eqnarray}
\widetilde{U}_{ii}^{(\ell)}({\bm q},z) & = & \int {\rm d}^2r_{i\parallel}~e^{i{\bm q}\cdot{\bm r_{i\parallel}}}U_{ii}^{(\ell)}({\bm r_{i\parallel}},z)\\
\nonumber
& = & \frac{e}{2\varepsilon_0\varepsilon_{{\rm eff}i}^{(\ell)}{\mathcal A}}~\int_0^{\infty} \frac{r_{i\parallel}J_0(qr_{i\parallel})}{\sqrt{r_{i\parallel}^2 + (z-\gamma_{\ell i})^2}}~{\rm d}r_{i\parallel},
\end{eqnarray}

\noindent where the effective relative static permittivities are defined as: $\varepsilon_{{\rm eff}i}^{(0)} = \varepsilon_i$, $\varepsilon_{{\rm eff}i}^{(1)} = \varepsilon_i/(\alpha_{{\rm t}i}^{(1)} + \alpha_{{\rm r}i}^{(2)})$, and $\varepsilon_{{\rm eff}i}^{(2)} = \varepsilon_i/\alpha_{{\rm t}i}^{(2)}$. The evaluation of the above integral yields:

\begin{equation}
\widetilde{U}_{ii}^{(\ell)}({\bm q},z) = \frac{e}{2\varepsilon_0\varepsilon_{{\rm eff}i}^{(\ell)}{\mathcal A}} ~\frac{e^{-q|z-\gamma_{\ell i}|}}{q}.
\end{equation}

The averages over the thicknesses of the layers are defined as:

\begin{equation}\label{widthpot}
\widetilde{U}_{ii}^{(\ell)}({\bm q}) = \int_{a_i}^{b_i} \psi^*(z)~
\widetilde{U}_{ii}^{(\ell)}({\bm q},z)~\psi(z)~{\rm d}z
\end{equation}

\noindent The lower and upper limits of the integral are defined as follows: if $i=1$, $a_1 = -3L$, and $b_1 = -L$; if $i=2$, $a_2 = L$, and $b_2 = 3L$. The evaluation of the integrals \eqref{widthpot} is easily achieved using the identities:

\begin{eqnarray}
&&\int \sin^2\left(\frac{\pi z}{w}\right)~e^{\pm qz}~{\rm dz} =\\
\nonumber
&& e^{\pm qz}\times\frac{\pm 4\pi^2 \pm q^2w^2 \mp q^2w^2\cos\left(\frac{2\pi z}{w}\right) - 2\pi qw \sin\left(\frac{2\pi z}{w}\right)}{8\pi^2q+2q^3w^2},
\end{eqnarray}

\noindent where $w$ is a nonzero real parameter. The following results are obtained:

\begin{equation}\label{uii0}
\widetilde{U}_{ii}^{(0)}({\bm q}) = f^{(0)}\times\frac{e}{2\varepsilon_0\varepsilon_{{\rm eff}i}^{(0)}{\mathcal A}} \frac{1}{q+\kappa_{{\rm eff}i}^{(0)}}
\end{equation}

\noindent with $f^{(0)} = \frac{\pi^2\left(1-e^{-qL}\right) -2q^2L^2e^{-qL}}{qL\left(\pi^2 + q^2L^2\right)}$;

\begin{equation}\label{uiil}
\widetilde{U}_{ii}^{(1)}({\bm q}) = f^{(1)}\times\frac{e}{2\varepsilon_0\varepsilon_{{\rm eff}i}^{(1)}{\mathcal A}} \frac{1}{q+\kappa_{{\rm eff}i}^{(1)}}
\end{equation}

\noindent with $f^{(1)} = \frac{e^{-2qL}\left(\pi^2+2q^2L^2\right)\sinh(qL)}{qL\left(\pi^2 + q^2L^2\right)}$; and 

\begin{equation}\label{uii2}
\widetilde{U}_{ii}^{(2)}({\bm q}) = f^{(2)}\times\frac{e}{2\varepsilon_0\varepsilon_{{\rm eff}i}^{(2)}{\mathcal A}} \frac{1}{q+\kappa_{{\rm eff}i}^{(2)}}
\end{equation}

\noindent with $f^{(2)} = \frac{e^{-4qL}\left(\pi^2+2q^2L^2\right)\sinh(qL)}{qL\left(\pi^2 + q^2L^2\right)}$. The screening parameters $\kappa_{{\rm eff}i}^{(\ell)} = f^{(\ell)}\times\kappa_{{\rm eff}i}$ account for the thicknesses of the layers and are computed in a consistent fashion from the three contributions $U_{ii}^{(\ell)}$, with $\ell=0,1,2$. For simplicity, $z_{01} = -2L$, and $z_{02}=2L$, and the intrawell electron-electron interactions $\widetilde{U}_{11}$ and $\widetilde{U}_{22}$ thus exactly have the same form, as can be expected.

The electron-electron interaction matrix element in the Fourier space finally reads:

\begin{equation}\label{veiei}
V_{ii}({\bm q}) = e\sum_{\ell=0}^2 \widetilde{U}_{ii}^{(\ell)}({\bm q})
\end{equation}

\noindent In the limit $L \longrightarrow 0$, one recovers the standard $q$-dependence of the Fourier transform of the statically screened 2D Coulomb potential, which, following the notations of Refs.~\cite{Laussy1,Laussy2}, reads:

\begin{equation}\label{std2d}
V_C({\bm q}) = \frac{e^2}{2\epsilon A}~\frac{1}{|\bm q|+\kappa}.
\end{equation}

\begin {figure}[!rh]
\centering
\scalebox{.325}{\rotatebox{0}{\includegraphics*{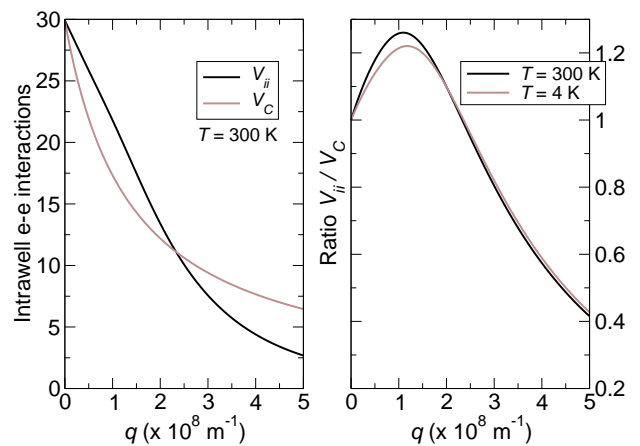}}}
\caption{Evidence of qualitative and quantitative effects of image charges on screened electron-electron intrawell interactions in two temperature regimes. The amplitudes shown in the left panel are scaled to the square of the electron electric charge.}\label{eeipot1}
\end {figure}

In order to compare the intrawell screened electron-electron interactions obtained with the two different models, the interactions $V_{ii}$ and $V_C$, and their ratio $V_{ii}/V_C$, are displayed as functions of the wavenumber $q$ on Fig.~\ref{eeipot1}. The amplitudes of $V_{ii}$ and $V_C$ are of the same order of magnitude but $V_{ii}$ is greater than $V_C$ for small values of $q$. Both functions are monotonically decreasing, but $V_{ii}$ possesses an inflection point while $V_C$ does not: this explains why the ratio $V_{ii}/V_C$ is not a monotonic function of $q$. Also, the ratio of the interaction amplitudes decreases rather fast as the wavenumber $q$ increases; this is due to $q$-dependence of $V_{ii}$, involving the layers width $L$ because of the presence of the image charges. $V_C$ does not depend at all on $L$. The curves for the ratio $V_{ii}/V_C$ are computed for two temperatures and show that the temperature plays a minor role in the discrepancies between the models under consideration.

\subsection{Interwell electron-electron interaction}
\label{intpara}

Let us now turn to the calculation of the potential in the quantum well $j=$ 1 or 2, created by an electron of the quantum well $i\neq j$, with effective mass $m_i$, and located in $(r_{i\parallel};z_{0i})$. The problem is more involved than the previous one since it concerns the evaluation of the interaction between two screened charges that are part of two separate electron gases with possibly different properties. We thus have to account for the dielectric mismatch and the screening effects in a consistent fashion.

Recalling the short presentation of the image charge method in Section \ref{imgchrg}, we see that the unscreened potential created beyond the interface \eqref{pot2}, by the reflected image charge \eqref{qrefl}, can also be viewed as a potential in an effective medium with an effective dielectric constant $\varepsilon' = (\varepsilon_1+\varepsilon_2)/2$. In the present work, we study a system that contains a dielectric layer between the two quantum wells. A practical way to proceed with our model is to consider that the regions below and above the plane $z=0$, are each characterized by an effective dielectric constant, $\varepsilon_1' = (\varepsilon_1+\varepsilon_{\rm d})/2$~ and $\varepsilon_2' = (\varepsilon_2+\varepsilon_{\rm d})/2$, respectively. In this case, accounting for the dielectric layer between the two quantum wells, we define the following potentials for $i\neq j$:

\begin{equation}
U_{ij} = \frac{e}{4\pi\varepsilon_0\varepsilon_j'}~ \frac{\alpha_{{\rm r}i}^{(1)}}{r_i},
\end{equation}

\noindent where the relevant first order image parameter and the distances $r_i$ are given by

\begin{equation}\label{qr1st}
\alpha_{{\rm r}i}^{(1)}=\frac{2\varepsilon_j'}{\varepsilon_i' + \varepsilon_j'}
\end{equation}

\noindent and 

\begin{equation}
r_i = \sqrt{r_{i\parallel}^2 + (z-z_{0i})}.
\end{equation}

\noindent In the definition of $r_i$, $z\in [L;L+l_{\rm qw}]$ if $i=1$, and $z\in [-L-l_{\rm qw};-L]$ if $i=2$, with $l_{\rm qw}=2L$.

Neglecting the $z$-dependence of $U_{ij}$, its Fourier transform simply reads:

\begin{equation}
\widetilde{U}_{ij}(\bm q) = \frac{e}{2\varepsilon_0\varepsilon_{\rm eff}{\mathcal A}q},
\end{equation}

\noindent where $\varepsilon_{\rm eff} = (\varepsilon_1'+\varepsilon_2')/2$. It is obvious from the expression above that $\widetilde{U}_{12}(\bm q) = \widetilde{U}_{21}(\bm q)$; but if one accounts for the Coulomb screening in each separate well directly by introducing the two effective screening parameters $\kappa_{{\rm eff}1}$ and $\kappa_{{\rm eff}2}$ in $\widetilde{U}_{12}(\bm q)$ and $\widetilde{U}_{21}(\bm q)$ respectively, as done in Eq.~\eqref{vii}, the equality of the potentials is violated. A convenient way to circumvent this problem is to consider that, in the same way as the effective dielectric constant $\varepsilon_{\rm eff}$ is defined, $\varepsilon_{\rm eff}$, one may also construct an effective static dielectric function $\varepsilon(\bm q) = [\varepsilon_1(\bm q)+\varepsilon_2(\bm q)]/2$, which explicitely reads:

\begin{equation}\label{statdf}
\varepsilon(\bm q) = 1 + \frac{\kappa_{12}+\kappa_{21}}{2q} = 1 + \frac{\kappa_{\rm eff}}{q},
\end{equation}

\noindent where the screening parameters $\kappa_{ij}$ are defined as follows:

\begin{equation}\label{kappaij}
\kappa_{ij} = \frac{\displaystyle m_i e^2}{\displaystyle 2\pi\varepsilon_0\varepsilon_{\rm eff} \hbar^2}~ \left(1 - e^{-\hbar^2\beta\pi N_i/m_i}\right).
\end{equation}

The parameters $\kappa_{ij}$ contain the effects of the dielectric mismatch through the presence of $\varepsilon_{\rm eff}$ in their calculation. Therefore, the effective static dielectric function $\varepsilon(\bm q)$ describes the mutual screened interaction of a charge of an electron gas with a charge of a separate, distinct electron gas, accounting for the different background dielectric constants of each system. Note that in the limit $N_i\rightarrow 0$, the screening parameter $\kappa_{\rm eff}$ \eqref{statdf} does not (and should not) reduce to $\kappa_{{\rm eff}j}$ \eqref{kappa}. In the present case, we seek to evaluate the screening felt by a test charge in a quantum well induced by an electron gas located in a separate and distant quantum well; the problem studied in Section \ref{eeintsec} is indeed quite different.

Accounting for the well width, the expression for the interwell screened electron-electron interaction reads: 

\begin{equation}\label{ve1e2}
V_{ij}({\bm q}) = f^{(2)}\times\frac{e^2}{2\varepsilon_0\varepsilon_{{\rm eff}}{\mathcal A}}~\frac{1}{q+f^{(2)}\kappa_{\rm eff}},
\end{equation}

\noindent where $\kappa_{\rm eff}$ is obtained from Eqs. \eqref{statdf} and \eqref{kappaij}, and the form factor $f^{(2)}$ is defined just below equation \eqref{uii2}. Again, in the limit $L \longrightarrow 0$, the standard $q$-dependence of the Fourier transform of the statically screened 2D Coulomb potential, is recovered.

In Ref.~\cite{Glazov3}, for a system geometry similar as that considered in the present work, the proposed form for the interwell screened electron-electron interaction reads,

\begin{equation}\label{gla}
V_{q} = \frac{2\pi e^2}{\kappa}~\frac{qe^{-qL}}{(q+q_s)^2 - q_s^2 e^{-2qL}}.
\end{equation}

\noindent Here, it is important to avoid confusion with the notations used in Ref.~\cite{Glazov3}: in Eq.~\eqref{gla} (and only in this case), $\kappa$ is the background dielectric constant and $q_s$, the screening parameter [also given by equation \eqref{kappa0}].

\begin {figure}[!rh]
\centering
\scalebox{.325}{\rotatebox{0}{\includegraphics*{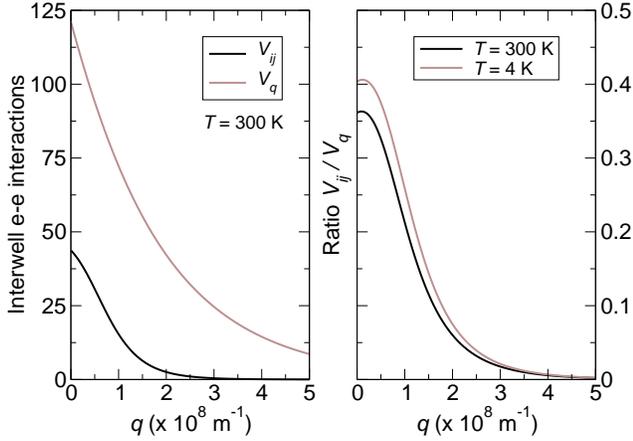}}}
\caption{Evidence of qualitative and quantitative effects of image charges on screened electron-electron interwell interactions in two temperature regimes. The amplitudes shown in the left panel are scaled to the square of the electron electric charge.}\label{eeipot2}
\end {figure}

The model interwell screened electron-electron interactions $V_{ij}$ and $V_q$ are represented as functions of the wavenumber $q$ on Fig.~\ref{eeipot2}. Both amplitudes are of the same order of magnitude, but unlike for the intrawell interactions, they do not cross each other: $V_{ij}$ is smaller than $V_q$ for all $q$. This shows that neglecting the polarization effects induced by the dielectric mismatch in the layered system, yields an overestimation of the interwell screened electron-electron interaction, which becomes increasingly important with $q$. The complicated $qL$-dependence of $V_{ij}$ yields an inflexion point which explains the non-monotonic behavior of the ratio $V_{ij}/V_q$. The curves are computed for two temperatures and , as for the intrawell interactions, the temperature plays a minor role in the discrepancies between the models. It is also interesting to note that the interwell amplitudes are greater than the intrawell ones: enchanced screening due to the dielectric mismatch lowers the strength of the effective Coulomb interaction between electrons of the same well in a more important fashion than for the interaction between electrons of differents wells; this reflects the non trivial effects of the form factors which not only tend to rapidly lower the range of the interactions, but also reduce screening.

\section{Interwell electron-exciton interaction}

Here, the system under consideration is characterized as follows: the quantum well 1 contains an electron gas of density $N_1$, and the quantum well 2 contains a dilute exciton gas of density $N_{\rm X}$; both gases are assumed to be in quasi-equilibrium at temperature $T$. Excitonic screening is neglected: only the free carrier gas contributes to Coulomb screening in the present model. Note that the density $N_2$ defined in the previous section is different from $N_{\rm X}$: $N_2$ corresponds to the density of the free carrier gas in the quantum well 2, and in this section we assume that $N_2=0$. 

The screened electron-exciton interaction in this system is calculated in a fashion similar to that developped in Section \ref{intpara}. The unscreened electrostatic potential experienced in the quantum well 2, generated by an electron located in $(r_{\parallel};z_0)$ in the quantum well 1 can be written as:

\begin{equation}
U_{12} = \frac{e}{4\pi\varepsilon_0\varepsilon_{\rm eff}}~\frac{1}{\sqrt{r_{\parallel}^2 + (z-z_{01})^2}},
\end{equation}

\noindent where $z \in [L;L+l_{\rm qw}]$ and $z_{01} \in [-L-l_{\rm qw};-L]$, with $l_{\rm qw}=2L$. The definition of $\varepsilon_{\rm eff}$ is the same as that given in Section \ref{intpara}.

Adopting the same procedure as in Section \ref{intpara}, yields the following expression for the screened interaction between an electron of the quantum well 1 and each of the bound carriers of an exciton in the quantum well 2:

\begin{equation}\label{ve1X2}
V_{12}({\bm q})= f^{(2)}\times\frac{e^2}{2\varepsilon_0\varepsilon_{{\rm eff}}{\mathcal A}}~\frac{1}{q+f^{(2)}\kappa_{\rm eff}}
\end{equation}

\noindent for the repulsive electron-electron interaction; an overall $-$ sign has to be included in the above formula for the attractive electron-hole interaction. Note the similarity between the expressions \eqref{ve1X2} and \eqref{ve1e2}. The essential difference comes from the nature of the gases in the two quantum wells and hence the screening of the interaction. The screening in the quantum well 2 is given by \eqref{kappaij} with $N_2=0$. The effective mass that enters the definition of the screening parameter in the case of this electron-exciton interaction is that of the electron of the quantum well 1. Note that if the effective masse of the electron in the quantum well 1 is different from that of the bound electron in the quantum well 2, the electron-exciton scattering matrix element contains only a direct term, which takes the following form \cite{Ramon,Ouerdane2}:

\begin{equation}\label{eXint}
V_{\rm eX} = V_{12}(\bm q) \times \left[\tilde{\psi^2}(\beta_h \bm q) - \tilde{\psi^2}(\beta_e \bm q)\right],
\end{equation}

\noindent where $\tilde{\psi^2}(\bm q)$ is the Fourier transform of the square of the wave function representing the relative motion of the bound electron-hole pair, and the parameters $\beta_{\rm e}$ and $\beta_{\rm h}$ are given by $\beta_{\rm c} = m_{\rm c}/(m_{\rm e}+m_{\rm h})$. If the quantum wells 1 and 2 are made of the same material, they are assumed to be sufficiently distant so that the overlap of the electon wave functions is zero; in this case, fermion exchange effects can be safely discarded and Eq.~\eqref{eXint} remains valid. 

To compare the above result with those obtained by the authors of Refs.~\cite{Laussy1,Laussy2}, it is useful to make some comments. First, as explained by the authors themselves in Ref.~\cite{Laussy2}, equation (2) of Ref.~\cite{Laussy1} is incorrect. Within the model~\cite{Laussy1} proposed by these authors, and keeping their notations, a careful derivation yields the following expression:

\begin{eqnarray}\label{VX}
&&V_X(\bm q) = \frac{32 e^2}{{\mathcal A}\epsilon_0\epsilon_{\rm r}a_{\rm B}^3}~\frac{e^{-qL/2}}{q +\kappa e^{-qL/2}}\\
\nonumber
&\times & \left[\frac{1}{\beta_{\rm h}^3\left( q^2 + 16/\beta_{\rm h}^2a_{\rm B}^2\right)^{3/2}} - \frac{1}{\beta_{\rm e}^3\left( q^2 + 16/\beta_{\rm e}^2a_{\rm B}^2\right)^{3/2}}\right]
\end{eqnarray}

\noindent where $a_{\rm B}$ is the 3D excitonic Bohr radius. Second, application of an external electric field perpendicular to the plane of a semiconductor quantum well induces a finite exciton dipole moment, which enhances and may keep finite the interaction of the excitons with the electron gas in the limit $q\rightarrow 0$ according to Ref.~\cite{Laussy2}. The response to an external electric field of the \emph{whole} system considered here is beyond the scope of the present work. Therefore, only Eqs. \eqref{eXint} and \eqref{VX} are compared. The exciton wave function entering \eqref{eXint} has, in principle, to be computed taking into account the screened electron-hole interaction \cite{Portnoi2,Ouerdane2}, and the exciton image charge effect as shown by Keldysh \cite{Keldysh1,Keldysh2}. Note that while screening and finite well width effects weaken the binding of the electron-hole pair, the image charge effect, on the contrary, enhances it \cite{Keldysh1,Keldysh2,Kumagai,Cen,Pauc}.

\begin {figure}[!rh]
\centering
\scalebox{.325}{\rotatebox{0}{\includegraphics*{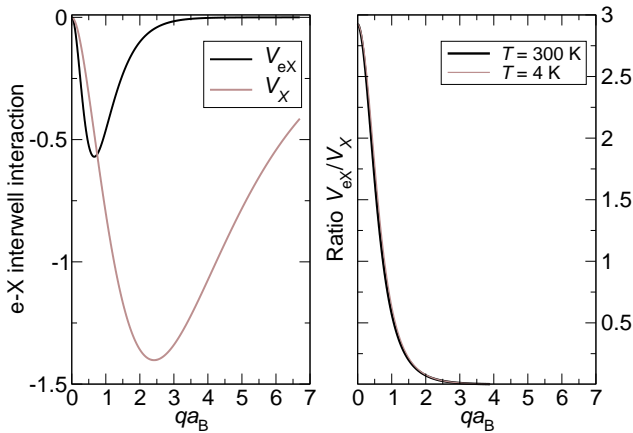}}}
\caption{Evidence of image charge effects on electron-exciton scattering matrix elements. The amplitudes shown in the left panel are scaled to the square of the electron electric charge.}\label{vexcomp}
\end {figure}

In Fig.~\ref{vexcomp}, the excitonic Bohr radius is calculated using the effective dielectric constant $\varepsilon_2'$ defined in the above section. Therefore, its value is $a_{\rm B} = 6.70$ nm (to be compared to $a_{\rm B} = 12.50$ nm with the dielectric constant of the sole GaAs-based quantum well). The discrepancy between $V_{\rm eX}$ and $V_X$ is important even though they present a rather similar shape: in the vicinity of $q=0$, the electron-exciton scattering matrix element that accounts for the dielectric mismatch, $V_{\rm eX}$, is almost three times greater than the amplitude obtained with the standard formula, $V_X$. The situation is reversed as the wavenumber $q$ increases; again this is due to the stronger $qL$-dependence of $V_{\rm eX}$, which tends to lower very rapidly its amplitude. The amplitude of the electron-exciton scattering matrix element computed with the standard formula also tends to be overestimated if the dielectric mismatch in the layered structure is not accounted for.

\section{Discussion}
The proposed model of screened electron-electron and electron-exciton interactions in layered structures accounts for polarization effects induced by the dielectric mismatch at the layers interfaces. It is based on a number of approximations but it yields useful analytical formulas for intrawell and interwell screened interactions, which can easily be implemented in a computer code. The formulas may be employed either to develop simple models of phenomena involving screened Coulomb interactions or to study limiting cases of more involved calculations for realistic models.

One of the main (implicit) assumptions is that the layered system does not interact with the environment within which it is embedded; in other words, from the electrostatic viewpoint, it is as though the two semiconductor layers were semi-infinite. To some extent, this issue is dealt with with the introduction of the $z$-component of the single-electron wave functions confined within the widths of the quantum wells (quantum tunelling between wells thus is neglected). The model also is restricted to a symmetric structure in order to avoid complications which would arise with differing layer widths; indeed in such a case the location of some of the image charges could be beyond the structure and additional assumptions would be needed to keep the derivations on the analytical level. Furthermore, one may anticipate that such generalization would bring little as far as the amplitudes of the interactions and their dependence on the transferred wavevector are concerned.

In the present work, modulation-doped semiconductor quantum wells were considered assuming that the dielectric mismatches between the wells and barriers materials were small enough to neglect the effects of image charges in the barriers. Extension of the present model to cases involving either important barrier doping or dielectric quantum wells characterized by important dielectric mismatch, is in principle possible in the spirit of previously published works \cite{Kumagai,Cen}. These works were focused on the enhancement of the exciton binding energies and nonlinear optical properties of dielectric quantum wells, but did not account for Coulomb screening.

Finally, as mentionned in the Introduction, the scope of the present work is restricted to quasi-thermal equilibrium situations for which the static limit is often used as a useful approximation, though it yields an overestimation of the RPA Coulomb self energies. Use of the dynamically screened Coulomb interaction permits predictive simulations of the properties of structures which are more complex than idealized quantum wells considered in the present article. Dynamical Coulomb screening is an essential feature of microscopic models of non-equilibrium systems for which a proper account of electron-electron scattering is necessary to compute the carrier distributions. A comparison of static and dynamic screening models indeed shows that the former model yields a spurious divergence of the scattering rates in the limit of small momentum transfer, while no such divergence problem occurs with the latter model\cite{Kane}. Numerical implementations of such sophisticated models, based on the nonequilibrium Green's function formalism \cite{Pereira2,Pereira3}, may represent a difficult task but they yield results in good agreement with experimental data.

\section{Conclusion}
The model presented in this article yields static screening parameters which account for the image charges induced by dielectric mismatch and the geometry of the considered system through forms factors. Application of the analytical formulas obtained for the screened Coulomb interactions in the present work and comparison to recently published models evidences non-negligible discrepancies. If one considers a layered structure with very small dielectric mismatch, one should not expect important consequences on the published results and conclusions \cite{Laussy1,Glazov1}, but it can be seen as worthwhile to make use of the simple formulas derived in the present article for the modeling of structures, which exhibit strong dielectric mismatch.

\section*{Acknowledgments}
I am pleased to thank Prof. Alexey Kavokin for the suggestion of the problem treated in this paper; Prof. Ivan Shelykh for providing useful remarks during the early stage of the work; and Dr. Patrick Bogdanski for interesting and fruitful discussions. I acknowledge partial support of the Agence Nationale de la Recherche.


%
%

%



\end{document}